# Relative estimation of scattering noise and its utility to select radiation detector for Gamma CT scanner.

Mayank Goswami[#] and Kajal Kumari

Divyadrishti Imaging Laboratory, Department of Physics, IIT Roorkee, Roorkee, India
Department of Physics, IIT Roorkee, Roorkee, India

[#]mayank.goswami@ph.iitr.ac.in,

## Abstract

This study investigates two unavoidable noise factors: electronic noise and radiation scattering, associated with detectors and their electronics. This study proposes a novel methodology to estimate electronic and scattering noise separately. It utilizes mathematical tools, namely, Kanpur theorem-1, standard deviation, similarity dice coefficient parameters, and experimental Computerized Tomography technique. Four types of gamma detectors: CsI (Tl), $LaBr_3(Ce)$, NaI (Tl) and HPGe are used with their respective electronics. A detector having integrated circuit electronics is shown to impart significantly less (~ 33% less) electronic noise in data as compared to detectors with distributed electronics. Kanpur Theorem-1 signature is proposed as a scattering error estimate. An empirical expression is developed showing that scattering noise depends strongly on mass attenuation coefficients of detector crystal material and weakly on their active area. The difference between predicted and estimated relative scattering is 14.6%.

The methodology presented in this study will assist the related industry in selecting the appropriate detector of optimal diameter, thickness, material composition, and hardware as per requirement.

## 1. Introduction

Computed tomography (CT) is an established non-destructive evaluation tool. Artifact-free CT images are desirable for obtaining true information about the scanned object. The presence of noise in the projection data of CT reduces the accuracy. The selection of the detectors plays an important role. Detectors are characterized based on their generated response function. The response function, commonly referred to as the energy spectrum, is the distribution of counts corresponding to different energies or pulse height amplitude. The response function of the detector depends on the size, shape, and composition of the detector's material[1]. The precise and accurate extraction of the energy spectrum can enhance the accuracy[2]. Transmission CT modeling must only include photopeak counts. However, the separation of counts due to Compton scattering from photopeak is not possible during the CT measurements.

The detector and its associated electronics add radiation counts to measured data due to two undesirable factors: (1) electronic noise and (2) the scattering of radiation with the detector's crystal material. Shrewd choice (about crystal material, its size, integrated onboard electronics, etc.) detector may minimize their effect.

The impact of electronic noise on the reconstructed image (referred to as image noise) is generally estimated using the standard deviation of CT numbers/pixel values (SDCT)[3]. Standard deviation (SD) of pixel values of reconstructed CT image in a uniform region of interest (ROI) is used. Two to three ROIs are selected in a single reconstructed CT image. The mean value of SD of ROIs is declared as SDCT.

The exact quantification of reconstruction error is only possible if prior information about the scanned object is available. If such information is already available would render the non-destructive evaluation unnecessary. However, Kanpur theorem-1 (KT-1) estimates the inherent error/noise level in projection data without having prior information about the scanned object[4]. The inherent noise involves both undesirable factors collectively. The following two subsections discuss the primary sources of the above-mentioned undesirable factors in detail:

### 1.1. Electronic nose

Electronic noise is primarily caused by: (a) noise from the preamplifier and integrating amplifier due to higher feedback capacitance: and (b) quantization noise caused by an analog-to-digital converter (ADC) over a broad signal range[5]. The electronic noise becomes comparable to the signal if a relatively weak (low activity) radiation source is used. It corrupts the CT examination's diagnostic value. Weak sources are required to reduce the probability of passive cancer and vascular disease development[6–8] due to radiation exposure. Noise reduction techniques are employed to minimize further and use low-activity sources to prevent electronic noise from distorting the signal. The noise reduction techniques use (a) statistically rich data measurement methods, (b) a digital filter-based data post-processing approach, (c) modification of underlying models, and (d) advanced hardware.

Measurement data with the best Gaussian fit, least standard deviation (SD), and low value of full-width half maxima (FWHM) is preferred and shown to give the best reconstruction quality[9]. Once the CT data is taken, post-processing noise filtering approaches may be applied during three stages: (a) before reconstruction[5,10], (b) while solving the inverse problem[11], and/or (c) after reconstruction on the pre-final image[12,13]. Several modifications are proposed in beer-lambert law for variance-related noise reduction to improve the CT Modelling[14].

As far as using advanced hardware is concerned, detectors and associated electronics are major components. Several comparison studies recommend detectors equipped with Integrated Circuit (IC) electronics over detectors equipped with distributed electronics to reduce electronic noise[3,15,16]. It was reported that image noise (arising due to electronic noise) could be reduced by approximately 10% (estimated by SDCT) using the IC detector[3]. The CT experiment was performed using a detector equipped with distributed electronics. The same CT experiment was performed using an IC detector, but this time, lowering the radiation dose. It was shown that a dose reduction of 20% could be achieved by replacing the detector equipped with distributed electronics with the IC detector. In both cases, SDCT values were matched[3]. This study claimed that the use of IC detectors reduces electronic noise, leading to further dose reduction.

### 1.2. Nonlinear Scattering of Radiation

Gamma radiation emerging from a nucleus may interact with atoms (in their mean free path) present in the rest of the material of the radioactive source and lose some energy, thus contributing to scattering part of the energy spectrum. The scattered and photopeak gamma radiation emitted from the radiation source further interacts with the object (under CT investigation) via three major phenomena: (1) Photoelectric effect, (2) Compton scattering, and (3) Pair-production. As listed above, the radiation passing through the object may interact with the detectors' crystal via three mechanisms. The whole phenomenon is illustrated in fig.1 [1]. The detector's response function thus may contain the scattering components due to scattering of radiation: (a) within the radioactive source itself, (b) inside the object (under CT investigation), and (c) inside the crystal of the detector. These components add a critical source of noise in the projection data (w.r.t. transmission CT) due to its inability to estimate and remove it in CT modeling. Radiation scattering affects the measured data quantitatively and qualitatively[17,18]. Because of the above-discussed issues, one may consider the scattering counts as an undesirable factor in transmission CT measurements. Several models and methodologies are developed to suppress or estimate and eliminate the scattering of radiation with the object from the measured data[19–27]. Scattering correction models aren't commonly employed in gamma transmission CT, but they're popular in single-photon emission computed tomography (SPECT), positron emission tomography (PET), and X-ray computed tomography. To the best of our knowledge, phenomena (a) and (c) are not investigated.

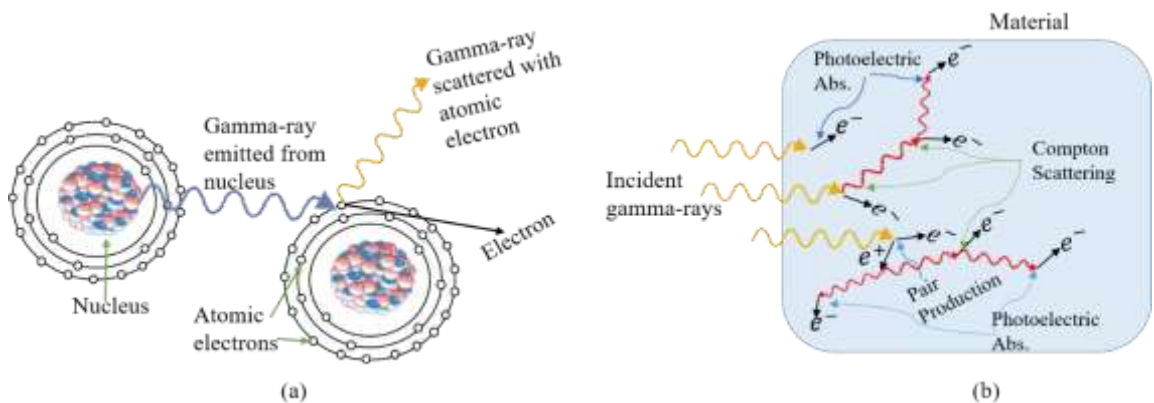

Fig.1: (a) Emergent of gamma rays from radioactive nuclei. The gamma radiation is scattered with atomic electrons and losses some energy, (b) The interaction of incident gamma radiation with matter through three major phenomena: (1) photoelectric effect, (2) Compton scattering, (3) pair production.

## 2. Motivation

Electronic noise is associated with detectors' electronics, but SDCT is estimated using an object (phantom or patient) profile. If an object with high-profile heterogeneity is used, then its SDCT may be relatively higher. It may or may not indicate that the detector and its electronics have higher electronic noise. Additionally, the variation of CT pixel values in uniform ROI may be due to ring artifacts. In this work, in section 3.1, the standard deviation parameter is used to estimate the electronic noise; however, an alternate approach is proposed.

To the best of our understanding, a tool to decouple and quantify scattering and electronic noise is missing in the literature. This work presents a methodology for the same. KT-1 estimates the relative scattering noise in projection data obtained by four gamma detectors.

Section 3 discusses material and methods; section 4 explains the results and discussion.

## 3. Materials & Methods

A compact gamma CT system is utilized as a tool to realize and evaluate the above-mentioned motivations. Four different types of gamma detectors are used in this work. The following section explains the implementation of error estimates.

### 3.1. Electronics Noise estimation methodology

A gamma radiation detector is placed in front of a Caesium source at 42cm (same as in the case of CT) but without having any object in between. Radiation counts are measured multiple times. Details of electronics settings are the same as set in CT measurement except window setting. The full-spectrum energy range is set in this case. It is estimated that the gamma source emission has insignificant energy-dependent randomness (maximum SD 2) as compared to randomness due to electronic noise. This calculation is explained in the supplementary file. Four different detectors having different crystal materials, sizes, and thicknesses are used. Scattering inside a given detector's crystal is dependent on its material distribution, sizes, and thicknesses. It is expected that scattering radiation contributes insignificant randomness in multiple measurement data. Since a single radioactive source is used, it is expected that the scattering of photons inside the radioactive source ($ScatN_{source}$) remain same for all measurements. The value of standard deviation in total counts is expected to give an estimate of electronic noise using this procedure.

### 3.2. Methodology to estimate relative error due to scattering

We cannot differentiate between photopeak counts and counts due to scattering (inside source, object, and detector crystal) in gamma photopeak. Kanpur theorem-I, however, is used to estimate the level of inherent error in the projection data sets of CT[28]. The detail of inherent error estimation is given in the supplementary file. This inherent error includes unavoidable electronic noise (EN), system noise due to suboptimal electronic setting (SysN), noise due to statistical inaccuracy (StatsN), reconstruction error (ReconN), and scattering noise (ScatN) due to scattering of radiation inside the material of radioactive source ($ScatN_{source}$), object ($ScatN_{sample}$), detector crystal ($ScatN_{crystal}$), and adjacent detectors and shielding ($ScatN_{other}$). It is assumed that these noise components are independent to each other thus are additive in nature as stated in Eq. 1a.

All four detectors are set at their respective optimal electronic settings that are obtained by statistically rich data measurement methods[9], except high voltage (HV). The value of HV applied to the PMT of each detector is set as received from the manufacturer, assuming these are optimal. Therefore, it is assumed that SysN has a negligible contribution to the inherent noise level. Error due to the reconstruction method (ReconN) is expected to impart a constant effect throughout the inverse image processing as the same reconstruction method is used. It converts the Eq. 1a into its approximate version Eq. 1b. Scattering component $ScatN_{other}$ can be nullified by not using the shielding and using only a single detector. The contribution due to scattering inside the radiation source $ScatN_{source}$ and object (being CT scanned) $ScatN_{object}$ will remain constant if the same source and object are used in every CT measurement with the same geometry. Since the thickness of the radiation source is negligible as compared to the object and detector crystal. So $ScatN_{source}$ impart negligible role in SysN. It basically transforms Eq. 2a into Eq. 2b for our studies.

$$\text{Inherent Noise} = EN + SysN + StatsN + ReconN + ScatN \quad \text{1a}$$

$$\text{Relative Inherent Noise} = EN + ScatN \quad \text{1b}$$

$$ScatN = ScatN_{source} + ScatN_{other} + ScatN_{object} + ScatN_{crystal} \quad \text{2a}$$

$$ScatN = ScatN_{object} + ScatN_{crystal} \quad \text{2b}$$

$$ScatN_{full\_spectrum} = ScatN_{compton} + ScatN_{photopeak} \quad \text{3}$$

Before utilizing KT-1 as a tool to evaluate the *relative* scattering noise (relative to the type of the used detectors) in this work, the sensitivity of KT-1 to noise due to scattering radiations is investigated. The following two strategies are employed to achieve this:

A. The gamma photopeak region contains fewer scattered photons as compared to the Compton region, while the full energy spectrum contains maximum scattered photons. In this strategy -

A, all factors (EN, SysN, StatsN, ReconN) which contribute to the inherent noise are kept constant except the ScatN. The scattering-related component (Eq. 2b) can be further defined based on the selection of the range of energy spectrum. It is expected that the maximum effect of ScatN $_{full\_spectrum}$ can be measured by selecting full spectrum (Eq. 3) settings during the CT measurement. The minimum effect of ScatN can be measured by selecting only the photopeak range. It can be verified using KT-1.

B. Two detectors of the same crystal material but different sizes are studied using KT-1. Secondary gamma radiation is produced when incident gamma radiation undergoes the Compton and pair production interactions. In small-size detectors, these secondary gamma radiations have more probability of escaping as compared to relatively larger or intermediate-sized detectors [1]. The increase in crystal size raises the photo fraction, owing to a higher contribution from Compton scattering as well[29]. The photo fraction is a direct measure of the probability that gamma radiation interacted with the material through any interactions, ultimately depositing all its energy within the detector. $LaBr_3(Ce)$ detectors of two different sizes $1'' \times 1''$ and $2'' \times 2''$ are used in strategy-B for CT measurements. KT-1 is applied to estimate ScatN $_{photopeak}$ because only photopeak range is selected. The noise components (referred in Eq. 1a) that contribute to the inherent noise are kept constant except the size of the detector's crystal by adopting strategy-B. Now, it is expected that if KT-1 is sensitive to scattering noise, then the projection data obtained from larger crystal size will show a relatively higher inherent noise level.

The details, specifications, and settings of electronic parameters of gamma detectors used in this study are given in the next subsection, 3.2.1. The detail of CT geometry parameters and phantom used in gamma CT is given in subsections 3.2.2. CT image reconstruction is provided in the last subsection.

### 3.2.1. Detectors and electronic settings

Four radiation detectors: a compact CsI (Tl) Multi Pixel Photon Counter scintillation detector with IC and three detectors with distributed electronics: (a) a high-resolution HPGe semiconductor detector, (b) a NaI (Tl) scintillation detector and (c) a $LaBr_3(Ce)$ scintillation detectors are used in this study. The specifications of these four different types of detectors are mentioned in table 1. Detectors with various crystal materials, sizes, and thicknesses are employed to explore the effects of these components. The latter set of hardware (make: Saint-Gobain Crystals) is interfaced with Nuclear Instrumentation Module (NIM) and multichannel data acquisition system. In NaI (Tl) detector (make: electronics enterprise Ltd. India), the scintillator crystal is coupled with its photomultiplier (PM) tube; Its anode output is amplified by the separate amplifier circuit. It is finally connected to a PC via a single-channel analyzer (SCA). The HPGe detector (make: Canberra and NI-PIXI-1031) is interfaced with NIM and a multichannel pixie-4 data acquisition system. IC electronics CT setup (make: Hamamatsu model C12137) contains CsI (Tl) scintillation crystal coupled with a photodiode, preamplifiers, and analog-to-digital converters; all are integrated onto the same silicon chip. Each of these has its own manufacturer-provided data acquisition software.

Table 1: Specifications of four gamma detectors

| Specifications of detectors | NaI (Tl) | CsI (Tl) | HPGe | $LaBr_3(Ce)$ |
|---|---|---|---|---|
| Crystal width (cm) | 5.1 | 1.3 | 5.6 | 2.54 |
| Crystal thickness (cm) | 5.1 | 1.3 | 6.0 | 2.54 |
| Hygroscopic | Yes | Slightly | Yes | Yes |
| Power Consumption* | +650V | +5V | -4500V | +643V |
| Estimated Resolution | 8.9% | 8.2% | 0.23% | 4.5% |

*Positive and negative sign shows the polarity of the bias voltage applied to detectors. HPGe semiconductor detector requires negative bias voltage, while NaI (Tl) and $LaBr_3(Ce)$ scintillation detectors require positive bias voltage. The CsI (Tl) detector requires a USB to get connected to a PC in plug-in-play mode.

The measured energy spectrum (count vs. energy plot) for all four detectors is shown in fig.2.

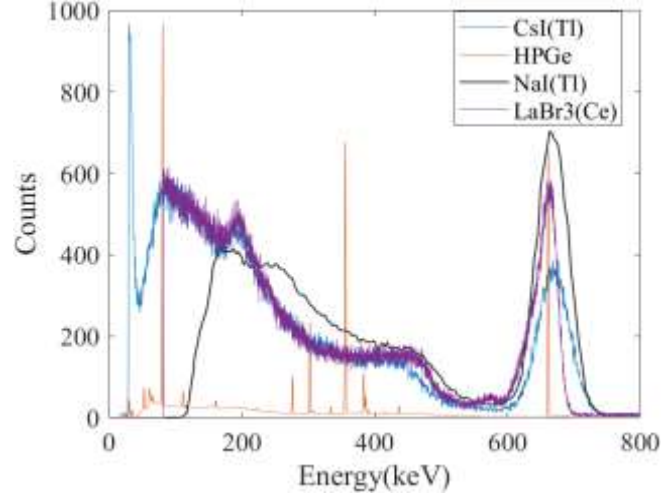

Fig.2: Energy spectrum of Cs-137 radioactive source detected by CsI (Tl), HPGe, NaI (Tl) and $LaBr_3(Ce)$ detectors.

### 3.2.2. CT Geometry and peripherals

The gamma CT experiment comprises an encapsulated radioactive source Cs-137 of activity 1.5µCi, an object, and gamma-ray detectors. For scanning purposes, a well-known cylindrical-shaped object/CT phantom is used. This object is made up of Perspex material with a diameter of 12 cm. The perspex also contains an aluminum cylinder (3.8 cm) in an off-centered location. Inside the aluminum cylinder, an iron cylinder (diameter of 0.8 cm) is inserted at the center location. It is shown in the inset location at the bottom right of fig. 3(c). The experimental setups are shown in fig. 3(a), 3(b) and 3(c). At a time, one type of gamma-ray detector is used to perform gamma CT experiments. The CT geometry is chosen such that enough counts are detected after being attenuated via the object. The geometry parameters are decided according to object size and source strength. Types of detectors have no role in deciding these parameters. Gamma CT experiment uses fan-beam geometry shown in fig. 3(d) with a fan-beam angle $\theta_{fan}$ of 37.1°. Source-to-detector and object-to-detector distances are kept at 42cm and 21cm, respectively. The object is uniformly rotated in 18 steps (each $\theta_{view}$ of 20°) between 0 to 360 degrees. This CT geometry is used for all CT experiments in this work. The full energy spectrum is recorded using a multichannel analyzer (MCA) for HPGe and $LaBr_3(Ce)$ detectors. CsI (Tl) detector has onboard/integrated MCA on its electronics chip. The counts for a particular energy range (as desired) are extracted from these data sets later. For NaI (Tl), desirable energy ranges are chosen using SCA.

During the experiment, the detector is translated on five locations, virtually building an array of 5 detectors in the shape of an 'arc' as shown in fig. 3d. The projection data matrix saved for each detector type is of size 5x18. This measurement scheme is referred to as Scheme 2a.

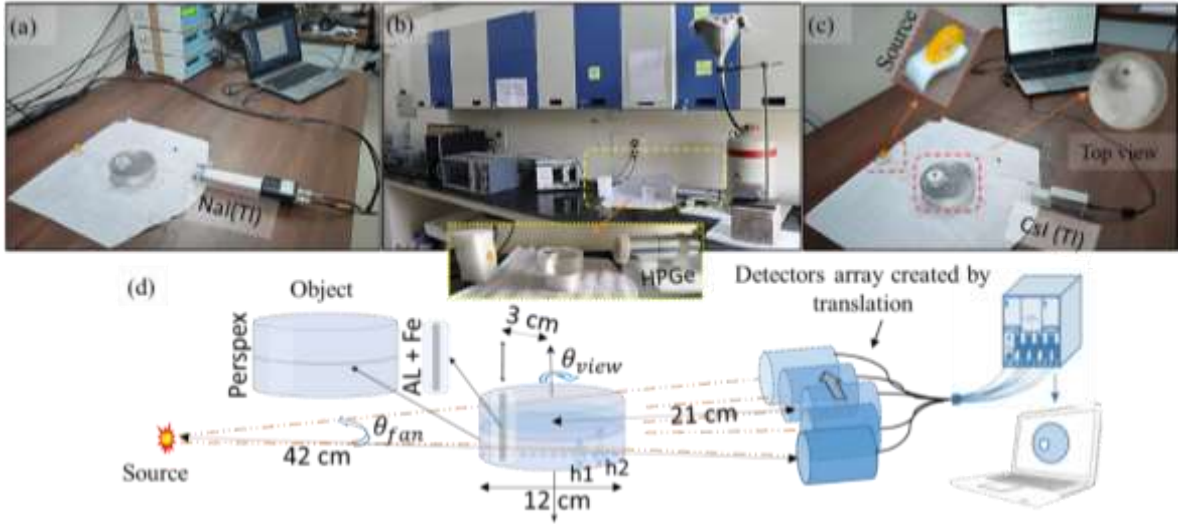

Fig.3: Gamma CT measurement Systems using three detectors: (a) NaI (Tl), (b) HPGe, (c) CsI (Tl), (d) Schematic diagram showing CT scanning geometry.

The 'arc' is formed in such a way that the first and last detector receives an equal spread of radiation passing via the object. It's not advisable to move the HPGe detector. Scheme 2b is devised to create four virtual detectors to complete the 'arc' of five and create CT data from 5 detectors for 18 views when the HPGe detector is used. The primary motive for using a single detector at a time is to avoid the $ScatN_{other}$.

We compared the sinogram taken by this *innovative* scanning method, scheme2b, with the scanning scheme2a and conventional scheme1 (when we would have used five detectors at the same time) for the other three types of detectors. It is found that all three scanning schemes give the same sinogram. The description of these three scanning schemes is given in the supplementary file.

### 3.3. Image reconstruction

KT-1 estimates the inherent noise in projection data if CT images are reconstructed using the convolution back-projection algorithm (CBP) for parallel beam geometry[30]. We first transformed the fan-beam projection data into a parallel beam equivalent to utilizing the Kanpur Theorem. Apt values of input information such as geometry information, number of detectors, number of rotations, and type of filter functions are utilized to develop custom reconstruction codes.

In our work, we didn't use the root mean square error (RMSE) as a reconstruction error (RMSE of reconstructed image w.r.t. cyber replica of object pixel to pixel) estimation tool. It is the least sensitive performance indices[31]. The Sorensen dice coefficient is employed to compute the similarity index in reconstructed images w.r.t. cyber replica[32].

## 4. Results and Discussion

The electronic noise estimation methodology (described in section 3.1) is applied, and the SD of four different detectors and their associated electronics (details mentioned in section 3.2.1) is evaluated, one at a time.

Figure 4 shows the SD of NaI (Tl), CsI (Tl), HPGe and $LaBr_3(Ce)$ $_{1''\times 1''}$ detectors. The lowest SD is observed for CsI (Tl) detector and the highest for the $LaBr_3(Ce)$ detector. However, a comparable SD is observed for NaI (Tl), HPGe and $LaBr_3(Ce)$ detectors. It shows that an IC-based detector, i.e., CsI (Tl) detector measurement has the least electronic noise as compared to other detector types equipped with distributed electronics. In IC detectors, an analog-to-digital converter is directly attached to the same board as the crystal material. This detector technology reduces the loss of information during the transfer of analog-to-digital signals. Apart from reducing electronic noise, IC reduces power consumption and heat dissipation[16]. The empirical rule of statistics is applied to all detectors to test the normality of data distributions. Three sigmas ($3\sigma$) rule is applied and 99.73% confidence interval is estimated for all the detectors. The calculation of $3\sigma$ limits is shown in supplementary data.

Strategy -A is adopted to perform CT experiments. Three different energy ranges: (a) gamma photopeak region, (b) Compton region (Compton edge + Compton valley), and (c) full energy spectrum, are selected, one by one. Three projection data: $P_{Photo}$, $P_{Comp}$ and $P_{Full}$ are obtained corresponding to these three energy ranges using a NaI (Tl) detector. The same procedure is followed for the CsI (Tl) and HPGe detectors. Due to the availability of two different sizes of $LaBr_3(Ce)$ detectors, we used these two for strategy B. However, strategy-A can be applied to these detectors also. This exercise has generated nine projection data sets, thus nine reconstructed images, in total. The noise level is estimated by applying KT-1 on reconstructed images. Root mean square error of goodness of fit ($GOF_{RMSE}$) is used as a KT-1 signature. It would be zero if noiseless projection data is used. The bar graph of $GOF_{RMSE}$ is shown in fig. 5. We infer that for NaI (Tl) detector, the $P_{Full}$ has maximum noise level and $P_{photo}$ has the least. The same is observed for the CsI (Tl) and HPGe detectors. Relative variation in noise, estimated for a single detector in three different energy ranges, would be due to scattering only as electronics and its settings (except spectrum range selection) are all same. Please refer to Eqs. 1b and 2b. Therefore, the inherent noise is basically due to ScatN (ScatN$_{object}$+ScatN$_{crystal}$). *This exercise validates the KT-1's sensitivity to noise due to the scattering of radiation.*

We cannot compare the scattering noise sensitivity of KT-1 signature between the type of detectors using figure 5 as described in Strategy - A. This can be accomplished by including all noise components referred to in Eq.1a. The major parameters that contribute to the variation of inherent noise are mainly due to electronic noise (EN) and ScatN, as discussed in section 3.2.

ScatN cannot be determined by simply subtracting inherent noise and EN values. It is further assumed that inherent noise value is made of individual weighted values i.e., $x$% and y% of EN and ScatN, respectively, are given in Eq. 4. The relative characteristics of EN (observed in fig. 4) are formulated in Eq. 5.

$$\text{Inherent noise} = x \text{ \% of EN} + y \text{ \% of ScatN} \qquad 4$$

$$EN_{CsI} < EN_{NaI} < EN_{HPGe} \qquad 5$$

Three possibilities arise while solving Eq. 4 and 5: (a) EN > ScatN, (b) EN ~= ScatN, and/or (c) EN < ScatN. The inherent noise, in all energy ranges, must follow the same pattern of inequalities as the dominant noise component follows. Possibilities of the case (a) and (b) cannot exist because, in the photopeak region, inherent noise follows the pattern $IN_{HPGe} > IN_{NaI} > IN_{CsI}$, which is not similar to EN (referring to fig. 5). The pattern of inherent noise is appropriate when the possibility of case (c) exists. It concludes that the scattering noise dominates over electronic noise irrespective of measurement range in the energy spectrum. Consequently, shows that the KT-1 signature is sensitive enough to estimate the relative scattering noise.

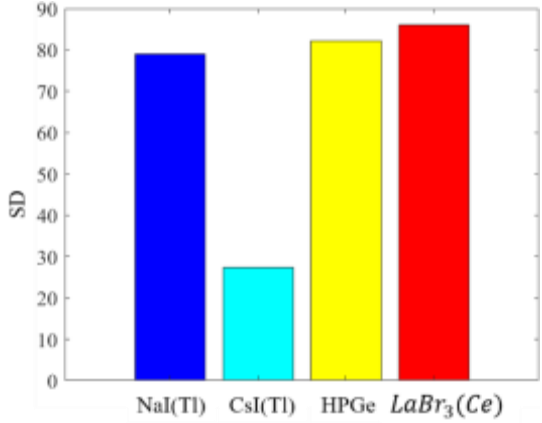 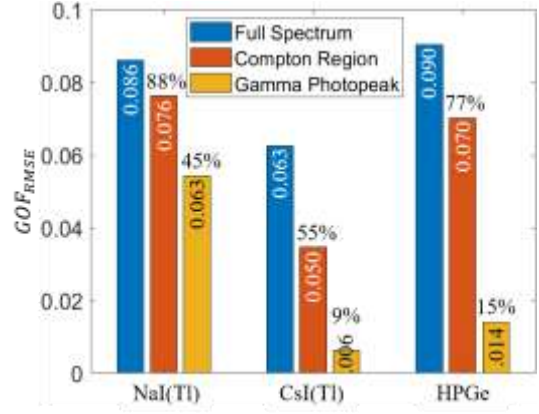

Fig. 4: Standard deviation of NaI (Tl), CsI (Tl), HPGe and $LaBr_3(Ce)$ detector.

Fig. 5: $GOF_{RMSE}$ as KT-1 signature of three different energy ranges obtained using NaI (Tl), CsI (Tl), and HPGe detectors.

$LaBr_3(Ce)$ detectors of two different sizes $1'' \times 1''$ and $2'' \times 2''$ are used individually to perform CT experiments using strategy B. The projection data is obtained by contributing the counts in the gamma photopeak region. The noise level in projection data is obtained using the KT-1 signature. The bar graph of $GOF_{RMSE}$ is shown in fig. 6. It is observed that projection data obtained from $LaBr_3(Ce)$ $_{1''\times1''}$ is less noisy as compared to those obtained from $LaBr_3(Ce)$ $_{2''\times2''}$. The variation in inherent noise is mainly due to scattering noise, as discussed in strategy B. The results obtained by KT-1 are in agreement with the theoretical statement mentioned in strategy B (an increase in the detector crystal's size raises the number of scattering radiations). This exercise validates KT-1's sensitivity to noise due to scattering radiation.

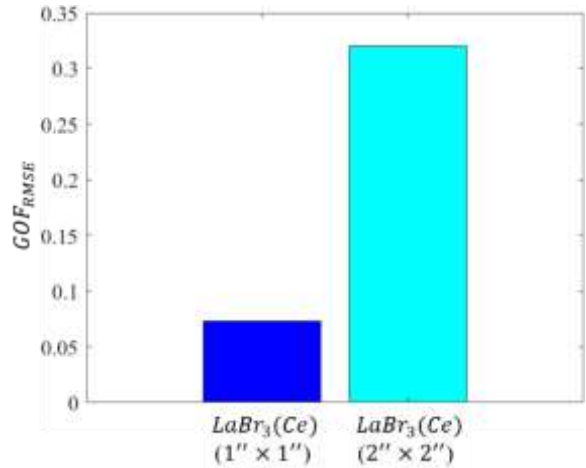

Fig. 6: $GOF_{RMSE}$ as KT-1 signature of $LaBr_3(Ce)$ detectors.

The scattering noise in four different gamma detectors is estimated using KT-1 once its sensitivity to scattering noise is confirmed. Gamma CT experiments are performed individually using CsI (Tl), $LaBr_3(Ce)$, NaI (Tl), and HPGe detectors. We have selected $LaBr_3(Ce)$ $_{1''\times1''}$ out of the two $LaBr_3(Ce)$ detectors because of their low scattering noise level. The bar graph of $GOF_{RMSE}$ as KT-1 signature of CsI (Tl), $LaBr_3(Ce)$, NaI (Tl) and HPGe detectors are plotted in fig.7, respectively. We note that the value of $GOF_{RMSE}$ is reflecting the value of scattering noise. It is observed that projection data obtained using CsI (Tl) detector has the least scattering noise while $LaBr_3(Ce)$ has the highest scattering noise.

The scattering noise depends on the active area of the detector crystal and the type of crystal material. The active area is calculated by the thickness of the detector crystal (as slice-wise in transverse plane CT modeling is performed) multiplied by its width. The details of the factors that affect the scattering are provided in table 2.

Table 2: The details of crystal thickness, diameter, and mass attenuation coefficient of four detectors

| Detectors | Crystal width (w) and thickness (t) | | The active area of detectors | Mass Attenuation Coefficient | Ref. |
|---|---|---|---|---|---|
| | w (cm) | t (cm) | $A_d$ ($cm^2$) | $\mu_m = \mu/\rho$ ($cm^2/g$) | |
| CsI(Tl) | 1.3 | 1.3 | 1.69 | 0.0700 | [33] |
| $LaBr_3(Ce)$ | 2.54 | 2.54 | 6.45 | 0.1028 | [34] |
| NaI(Tl) | 5.1 | 5.1 | 26.01 | 0.0758 | [35] |
| HPGe | 5.6 | 6.0 | 33.60 | 0.0745 | [36] |

The relation between scattering noise estimated using KT-1 signature, mass attenuation coefficients ($\mu_m$) of detector crystal material and their active area ($A_d$) is fitted in equation 6 with 2% of RMSE in fitting. It is shown in fig.8. The data given in Table 2 is also used. It is observed that the noise due to scattering (inside the object and detector crystal) will be larger if crystal material with a high mass attenuation coefficient with relatively thick crystal is used.

$$GOF_{RMSE} = 863.7(A_d^{0.3} \times \mu_m^{4.3}) \qquad 6$$

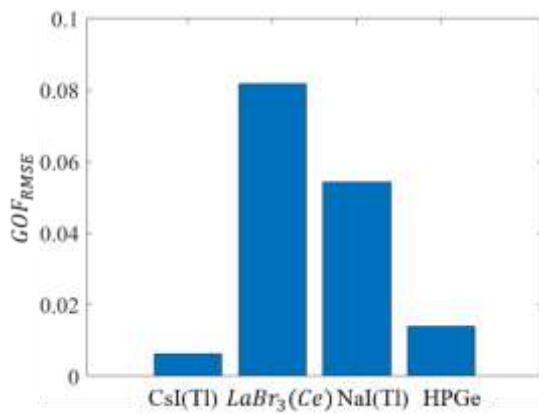
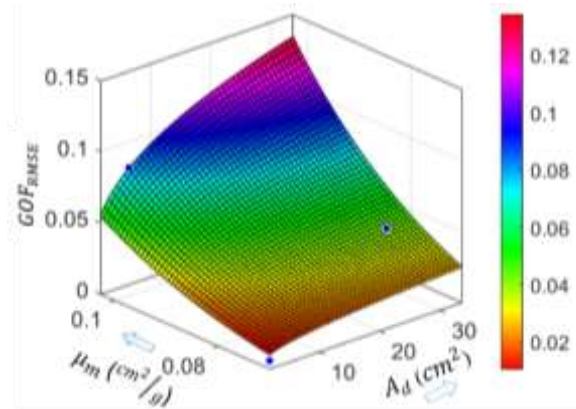

Fig. 7: $GOF_{RMSE}$ as KT-1 signature of CsI (Tl), $LaBr_3(Ce)$, NaI (Tl) and HPGe detector, respectively.

Fig. 8: $GOF_{RMSE}$ vs. Mass attenuation coefficient and active area of the detector.

A fresh NaI (Tl) detector that is not used to develop the empirical relation in Eq. 6 is used to measure another CT data using the same object and geometry. The value of $GOF_{RMSE}$ obtained is 0.041. Its crystal material has an attenuation coefficient 0.0758 $cm^2/g$ and its active area is 26.01 $cm^2$. Predicted GOF by Eq. 6 is 0.035. We note that this empirical relation is developed using the data of only four detectors having different materials, and sizes. The difference (14.6%) in GOFs by Eq. 6 and measured using CT may be reduced by carrying out the same study using many detectors of different types.

The reconstructed CT images obtained by four gamma detectors are shown in fig. 9. The value of dice similarity coefficients for reconstructed images w.r.t. cyber replica of the object is calculated and given below to the respective images. Figure 9(a) and 9(b) shows the cyber replica of the object (2-D image of a scanned object) and its simulated reconstructed CT image. The highest value of the similarity index is observed for CT simulated reconstruction of the cyber replica of the object (fig. 9(b)). We note this reconstruction presents a noiseless data reconstruction case without real measurement and that has zero scattering noise. Figure 9(c), 9(d), 9(e), and 9(f) shows CT images reconstructed using projection data obtained by CsI (Tl), $LaBr_3(Ce)$, NaI (Tl), and HPGe detectors, respectively. Reconstruction from CsI (Tl) detector scanner has the highest similarity index, while the HPGe detector scanner resulted lowest.

The process of CT measurement imbibes information about a pixel in the line integrals crisscrossing it according to (a) the intercept of lines inside that pixel and (b) the pixels' linear attenuation coefficient. The latter also depends on the energy of radiation crossing that pixel. The process of reconstruction redistributes this information back into that pixel using the same factors. The reconstruction algorithm will redistribute relatively less amount of this information back to a particular pixel having crossed vertically or horizontally as compared to the pixel which is crossed diagonally. Similarly, a pixel with high attenuating material distribution will be get redistributed more information back to it. Any error in redistribution in a particular pixel will reflect the wrong redistribution of the overall information to other pixels since the inverse process solves all relations simultaneously. Any dissimilarity between the reconstructed image of noiseless data (i.e., fig. 9(b)) with reconstructed images of real data can be attributed to noise (particularly error in modeling by not incorporating scattered radiation into line integrals) in the data itself. It is assumed that the reconstruction error will remain the same in each image.

We have quantified this similarity using the Dice coefficient (exact values are given in figure 9 as well) of: (a) the overall image ($D_1$) and (b) region containing perspex only ($D_2$). Figure 9(b), when compared with fig. 9(a), the only dissimilarity is visible in the metallic region. The reconstructed region of the perspex is approximately the same. The dice similarity coefficient of the overall image is ~0.8, and only of the Perspex region is ~0.7. These are the best that can be recovered. The dissimilarities in recovery images from the real-world measurement (fig. 9(c-f)) are highlighted. The displacement of wrongly distributed information is marked using orange-colored arrows with yellow markers. The excess and deficiency but wrongly distributed information is marked using yellow and red color boundaries of respective shapes. The qualitative and quantitative analysis coherently shows that the recovered image with such bigger regions has low dice coefficients, especially $D_2$. As these regions are due to scattering noise, they must be associated with $GOF_{RMSE}$ quantitatively as well. We also note that except for reconstruction from the scanner using the HPGe detector Dice coefficient is following the same trend, i.e., images recovered from CsI (Tl) have the least $D_1$, $D_2$ and $GOF_{RMSE}$ and so on. We infer that dice coefficients reflect the distribution of noise in the reconstructed image.

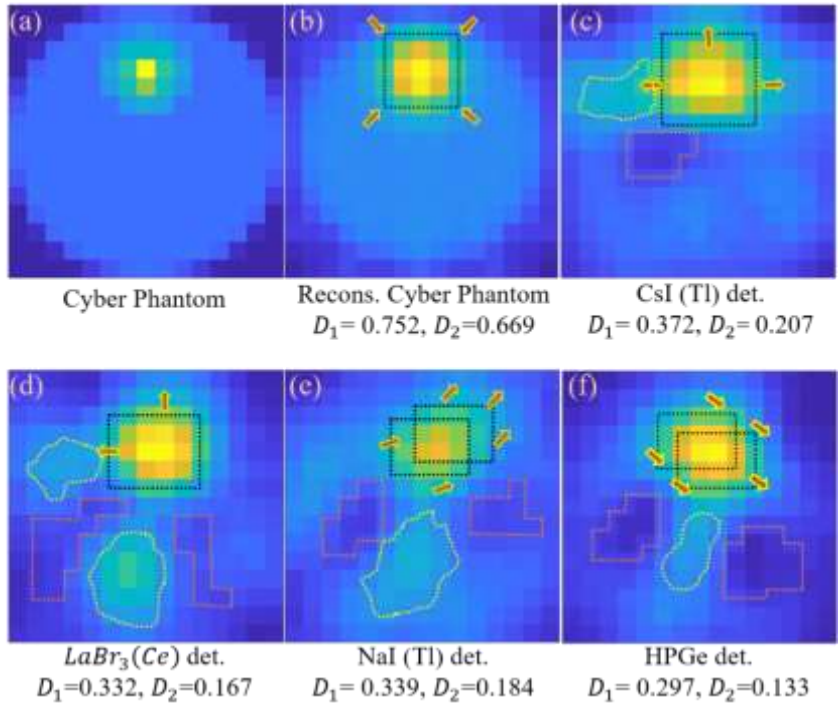

Fig.9: (a) Cyber replica of the object, CT Image obtained by (b) reconstruction from cyber replica data, (c) CsI (Tl), (d) $LaBr_3(Ce)$, (e) NaI (Tl), and (f) HPGe detector.

The experiments are performed by removing aluminum and iron cylinders as well to study the effect of the metal portion on scattering signatures. The results are given in the supplementary file. The scanned object is made up of 2 layers of perspex. These layers are combined with the help of glue. Some patches are observed in the object profile. It is confirmed through an experiment that the apparent patches of efficient and inadequate information in CT images are not the result of the glue layer given in the supplementary file. The CT images reflect the combined effect of noise due to scattering with the object and detector crystal material; however, the scattering noise depiction is all dumped on the object as CT modeling does not include cross-section of the crystal. HPGe crystal size is large enough to cover the heterogeneity in every translation. This may be the reason we have no congruence between dice values and $GOF_{RMSE}$.

## 5. Conclusion

An alternate approach to SDCT is proposed to estimate the electronic noise in detectors equipped with different types of electronics. Mainly two types of electronics (IC and distributed) are compared. It is observed that the detector equipped with IC has low electronic noise in data as compared to the detector equipped with distributed electronics.

This study presents a methodology utilizing transmission CT to estimate the noise due to the scattering of radiation with the detector's crystal and scanned object. The sensitivity of KT-1 to scattering noise is demonstrated for the first time, and it is used to estimate the relative scattering noise in a different types of gamma detectors. A relation between noise due to scattering radiation, the detector's crystal material, and the active area of the detectors is demonstrated. One must take into account the tradeoff parameters of detectors like material type, active area, noise level (electronic and scattering noise), power consumption, and heat dissipation to make an advanced CT system. The influence of scattering noise on CT images is also illustrated. This study also concludes that the dice similarity coefficient is less sensitive for the estimation of scattering noise than $GOF_{RMSE}$.

## Acknowledge

K. Kumari is thankful to the Council of Scientific and Industrial Research (CSIR), India, for the fellowship. M. Goswami would like to acknowledge Prof. Anil K. Gourishetty, Prof. Ajay Y. Deo, and members of the RDS Lab at the Department of Physics, IIT Roorkee, for allowing to use the HPGe and LaBr3 detector systems. M. Goswami wishes to thank Dr. Snehlata Shakya for sharing the grid-based CBP code.

## Credit authorship contribution statement

K. Kumari: Methodology, Experiments, Data Acquisition, coding, Data Analysis, and writing. M. Goswami: Methodology, Data Analysis, Funding, Supervision, and Writing.

## Conflicts of Interest: The authors declare no conflict of interest.